\documentclass[sigconf,nonacm]{acmart}

\usepackage[utf8]{inputenc}
\usepackage[english]{babel}
\usepackage{tikz}
\usepackage{pgf-pie}
\usepackage{xcolor}
\usepackage{longtable}

\usepackage{algorithm}
\usepackage{algorithmic}

\usepackage{listings}
\lstdefinelanguage{diff}{
    basicstyle=\ttfamily\scriptsize,
    morecomment=[f][\color{gray}]{@@},
    morecomment=[f][\color{green}]{+\ },
    morecomment=[f][\color{red}]{-\ },
}
\usepackage{subcaption}

\author{Adam Krafczyk}
\affiliation{
    \institution{University of Hildesheim}
    \department{Institute of Computer Science}
    \city{Hildesheim}
    \country{Germany}
}
\email{krafczyk@uni-hildesheim.de}

\author{Klaus Schmid}
\affiliation{
    \institution{University of Hildesheim}
    \department{Institute of Computer Science}
    \city{Hildesheim}
    \country{Germany}
}
\email{schmid@uni-hildesheim.de}

\acmConference[EASE 2026]{The 30th International Conference on Evaluation and Assessment in Software Engineering}{9–12 June, 2026}{Glasgow, Scotland, United Kingdom}

\title{Reproducible Automated Program Repair Is Hard -- Experiences With the Defects4J Dataset}
\thanks{This work was carried out within the ITEA 4 project GENIUS, as part of the ITEA programme, the Eureka Cluster on software innovation. This work was funded by the German Federal Ministry of Research, Technology, and Space (BMFTR) under grant number 16IS24069H}


\begin{document}

\begin{abstract}
In the research of automated program repair (APR), benchmark datasets consisting of known defects in combination with test suites that indicate the defects are of high importance.
They allow for an evidence-based comparison of different APR approaches.
In our own work on APR we found significant challenges when working with widely used defect datasets, which go beyond mere repeatability of defects via test cases.
We summarize these identified challenges and related lessons learned to bring them to the attention of the APR community and quantify the potential impact of them.

In particular, we investigate the widely used benchmark Defects4J, which has according to Google Scholar over 1,800 citations.
It consists of 835 defects from 17 open-source Java projects; a hand-curated collection of defects, test suites that clearly indicate the defect, and human patches where any unrelated changes are removed.
We find that, when executing the test suites with strict requirements for reproducibility in APR settings (beyond merely reproducing the defect via test cases), 180 (21.6\,\%) of the defects are not suitable for evaluation experiments.
Further, we find that an additional 59~(7.1\,\%) defects have test suites that are obviously under-specified, as deleting a single statement from the code base makes all test cases pass, although the human-written patch does not only delete code.

Our contributions are: a systematic collection of requirements for defect datasets for APR beyond traditional reproducibility of defects, a description of practical experiences and quantitative analysis of problems with the Defects4J dataset, as well as an implementation of an evaluation framework for APR tools for Java programs.
This evaluation framework does stricter checking for indications of inadequate test suites, to avoid otherwise unnoticed problems in the test suite, such as flaky tests.
\end{abstract}

\maketitle

\section{Introduction}
\label{sec:introduction}
In scientific evaluations of automated program repair (APR) approaches, benchmarks serve a crucial role.
They consist of implementations of systems, containing known defects, often from real-world projects, together with test cases and expected patches.
These datasets are used to compare different APR approaches by simply measuring how many defects each approach is able to fix.
Often consisting of open-source software, they help to make experiments in published work reproducible by other researchers.
Typically, defect datasets become quasi-standard in the evaluation of tools with the same repair scope (mostly programming language).
As many works that introduce new repair techniques will typically use the readily available, well established datasets to evaluate the capabilities of their approach, there directly emerges the possibility to compare different approaches.

One such well-established defect dataset for the Java programming language is Defects4J~\cite{JustJalaliErnst14}.
According to a survey of defect datasets, it is the most used benchmark for APR evaluations~\cite{ZhuFurthPradel+25}, making it a quasi-standard evaluation benchmark for APR tools targeting Java programs.
In its 2.0 version, it consists of 835 defects from 17 different open-source projects\footnote{\url{https://github.com/rjust/defects4j/tree/v2.0.0}}.
Defects4J provides openly available tooling to reproduce the buggy and fixed versions of each defect, as well as enabling compilation and test suite execution for repair attempts.
The defects in this dataset are manually isolated; unnecessary changes, that may have snuck into the original commit that fixed the defect, are removed so that only the relevant changes are recorded in the dataset.
This makes the comparison of automatically generated patches with the human-created original patch easy.

In our own work on APR for Java, we benefited immensely from Defects4J, as it is readily available and well-established in the research community.
However in practice, we identified a number of significant challenges when working with defect datasets such as this.
These are both of a  technical nature, as well as more nuanced potential problems, which only come to light under more thorough inspection.
Especially the not-so-obvious problems have the risk of significantly influencing the evaluation results when using the dataset.
They unfortunately also tend to go unnoticed, if not specifically looking for them.

We find that the requirements which defect datasets must fulfill to be suitable for evaluating APR approaches go beyond what is traditionally considered for reproducibility of defects.
While Defects4J is considered to have very high reproducibility~\cite{ZhuRubioGonzalez23}, investigations here focus on whether the defective project can be built, tests can be executed, and the expected failure can be observed.
In an APR context, however, further and stricter requirements arise, such as a consistent test suite results for partial and out-of-order executions of tests.
Thus, we define the \emph{workability} of defects in an APR context, which builds upon and extends the traditional notion of mere reproducibility of a defect.

In this paper, we report on our experiences when working with defect datasets for APR and suggest improvements over the status quo.
Section~\ref{sec:requirements} outlines general requirements for that defect datasets must fulfill to be considered \emph{workable} for scientific evaluation of APR techniques.
Section~\ref{sec:defects4j} then looks at the specific case of Defects4J and the degree to which it fulfills these.
We describe in Section~\ref{sec:environment} our environment for testing the reproducibility of defects.
This introduces  our implementation of an evaluation framework that strictly checks for indications of inadequate test suites in the underlying project.
Section~\ref{sec:problems} describes the actual problems we encountered working with the Defects4J dataset, while Section~\ref{sec:solutions} proposes (sometimes only partial) solutions for these problems.
Further, in Section~\ref{sec:test suite-quality} we dive further into an analysis of the quality of the test suites provided by the underlying open-source projects included in Defects4J, with respect to their suitability for APR tools.
In Section~\ref{sec:threats} we discuss the threats to validity for the reproducibility and test suite quality experiments.

\section{Related Work}
\label{sec:related-work}

Reproducibility in science has been significantly scrutinized over the last few decades, partially triggered by negative reproducibility results in psychology and sociology~\cite{Baker15}.
Also in the area of software engineering reproducibility received significant attention leading to different approaches for assessment and improvement of reproducibility in empirical studies~\cite{CordeiroOliveira25}.

The area of empirical research based on (publicly available) software repositories provides a rather good basis for reproducible research.
While initial research by González-Barahona and Robles found the status of research wrt.\ reproducibility lacking~\cite{GonzalezBarahonaRobles12} their more recent study found a significant improvement of the status of reproducibility~\cite{GonzalezBarahonaRobles23}.

The area of automated program repair (APR) significantly relies on the quality of datasets for reproducibility.
A very large number of such defect datasets have been created to analyze different questions around programming.
In their survey Zhu et al.\ were able to identify 132 such datasets~\cite{ZhuFurthPradel+25}.
Out of these one of the most important defect datasets is Defects4J~\cite{JustJalaliErnst14}.
Due to  its importance its reproducibility has been heavily scrutinized from several angles.

A reproducibility study of five defect datasets consisting of Java programs finds all defects in Defects4J version 2.0 to be reproducible~\cite{ZhuRubioGonzalez23}.
They use the Defects4J tooling to compile the programs and run the test suite to determine if the defect is exercised by the test suite as expected.
In contrast, in this paper we check for \emph{workability}, i.e.\ stricter in-depth checks in our evaluation framework that test additional criteria, such as the consistency of the test suite results, which reveals additional issues.

The flakiness of individual test methods in Defects4J version 1.3.0 has been investigated by Qin et al.~\cite{QinWangLiu+21}.
They analyzed the test methods that have been marked flaky by the Defects4J authors.
Note that the Defects4J tooling automatically removes all known flaky tests when checking out the projects, so experiments are not affected by these investigated flaky methods by default.
Additionally, they find that given a stable execution environment makes most of the flaky tests behave deterministic.
In this paper, we adhere to a stable environment as defined by the Defects4J authors and find additional non-determinism in the dataset, which goes beyond the scope of the previous study.

Different methods for detecting flaky tests have been proposed, such as looking for differences in the coverage data when executing the test multiple times~\cite{GruberFraser23}.
In this paper, we detect flakiness of test suites by observing behavior of the test suite and system under test that are relevant for typical APR tools (c.f.\ Section~\ref{sec:environment}).

Besides pure reproducibility issues Defects4J has been criticized also for its applicability for APR.
Martinez et al.\ found in an analysis of bug repairs that used Defects4J that while they found many repairs that made the test suites pass, the resulting programs were not correct, leading them to the conclusion that the tests given in Defects4J are critically underspecified~\cite{MartinezDurieuxSommerard+17}.
Rafi et al.\ observed that Defects4J defects contain tests in their test suite that were originally created together with the bugfix (same commit); this brings into question whether fault-localization techniques can be properly evaluated using Defects4J as the test suite (for some defects) contains developer knowledge of the fix~\cite{RafiChenChen+25}.
Also Sobreira et al.\ found that the bug fixes contained in Defects4J are rather small and often contained only code additions~\cite{SobreiraDurieuxMadeiral+18}.
Another issue for techniques that use modern LLM-based techniques is that Defects4J is part of their training set~\cite{ZhangZhangZhai+24, LeeKangYoon+24}.
While the aforementioned issues are important restrictions to the applicability of Defects4J in the APR context, it is not directly related to our study as we focus only the reproducibility of the individual defects themselves and do not discuss reproducibility for specific APR techniques.

\section{Requirements on Defect Datasets}
\label{sec:requirements}

A benchmark for automated program repair (APR) consists of a collection of known defects.
Each entry in the dataset has a  version of the source code as well as a test suite with at least one failing test case,  indicating a defect that needs to be fixed.
Typically, the defect dataset also contains the original, human-written fix for the defect; this is not supplied to the benchmarked APR tool, but can be used to compare the automatically created patch with the human-written one.

The defects in the defect dataset must have certain characteristics to be \emph{workable} for scientific evaluation of APR approaches.
In the following list, we introduce them as requirements that need to be fulfilled.
These range from trivially seeming technical aspects, which we still list for the sake of completeness, to more nuanced aspects that are nevertheless important for scientific rigor of the resulting experiments.
These requirements are based on our experiences with setting up evaluation environments for APR tools as well as on typical assumptions made by APR tools.
We use defects for Java programs as an example, but the requirements can easily be adapted for other programming languages.

\emph{Workability} of a defect for APR as defined here is an extension of the concept of defect reproducibility.
Reproducibility requires that the defective program can be built, the test suite can be executed, and (only) the expected failure is indicated by the test result.
Our \emph{workability} requirements further aim at additional qualities that need to be present in order for the defect to be usable for reliable evaluation of typical APR approaches.
However, as APR approaches differ significantly, not all requirements are strictly necessary for all APR approaches.
Thus, we indicate for each requirements the scope of affected APR approaches.

\begin{description}
    \item[R1]
    The source files of the program need to be defined and parseable; APR tools need to know which parts of the program can be modified and they need to be able to build a representation of the code by parsing it (or otherwise accessing them in an automated fashion).
    This requirement is necessary for all APR approaches, as they all need to modify the source code somehow to produce a fix.

    \item[R2]
    The project source files need to be able to be compiled.
    Ideally, this can be done by simply invoking a compiler on the (potentially modified) source files (potentially in a different (temporary) location).
    For instance for Java programs, the classpath with all dependencies needs to be specified (we call this the \textit{compilation classpath}).
    A build setup that is more complex than just invoking a compiler on the source files makes it harder for APR tools to create variants of the program for evaluation; a generic approach (simply invoking the compiler) is very much preferred, although not strictly necessary.
    This requirement is necessary for all APR approaches, as they all need to evaluate modifications to the source code.

    \item[R3]
    The project must have a test suite that can validate the program (and variants of it).
    For Java, the tests are typically written with the JUnit test framework.
    Executing them requires the location of the compiled program classes (possibly modified), the compiled test classes, and a classpath with any dependencies necessary for executing the tests (we call the latter two combined the \textit{test classpath}; the compiled program classes differs from execution to execution, as the different variants are evaluated; we expect the APR tool to add them to the test execution as required).
    This requirement is necessary for all APR approaches, as they all need to evaluate modifications to the source code.

    \item[R4]
    The test suite must report individual test case results.
    An aggregated ``passes'' or ``fails'' for the whole test suite, without the result of individual test cases, is not enough.
    This requirement is only necessary for APR approaches which use fitness functions or similar, for which they need a detailed list of which test cases are failing and which ones are passing.
    This is very common in practice, but there could exist APR tools which do not require detailed test case results.

    \item[R5]
    The test suite must allow individual test cases to be executed and allow coverage information to be collected.
    This is required as fault localization needs to have coverage information for all individual tests to compute suspiciousness values for individual statements.
    With JUnit, this is typically easily achievable by creating a small test driver that only executes a single test via the JUnit API.
    Coverage information can be collected by attaching a JaCoCo agent to the test process.
    This requirement is only necessary when fault-localization is based on the coverage information of test cases.

    \item[R6]
    The test suite must produce consistent results for the same variants.
    Flaky tests, which alter between failing and passing randomly for the same variant, greatly disturb any APR process, which is based on generate-and-validate.
    Furthermore, running individual test cases in any order or the whole test suite at once should always produce the same result for the same variant and test; or stated the other way round: test cases should not depend on order or other tests being executed.
    The test cases being non-flaky is required for all APR approaches, but out-of-order or individual execution of test cases only needs to be consistent when the APR tool executes them as such.
    For example, many APR tools use coverage-based techniques to only execute the necessary subset of tests to validate a certain program variant; the reasoning here is that test suite execution is often the main driver for the runtime of a repair approach, and thus reducing the number of tests per variant validation significantly reduces runtime.
\end{description}

\section{Experiences and Observations with Defects4J}
\label{sec:defects4j}

In this section we look at the specific example of Defects4J, one of the most widely used defect datasets for the task of automated program repair~\cite{ZhuFurthPradel+25}.
While this is only one dataset, the points made here can also be adapted and applied to other defect datasets.
The problems we find range from a technical nature to more fundamental ones; the fundamental problems are directly relevant for other defect datasets as well, while the technical problems are more specific to the particular case of Defects4J.
However, even the more technical problems may be relevant to other defect datasets as well, though they likely require adaptation to the particular setup of the datasets.

First, in Section~\ref{sec:environment} we describe our setup for testing \emph{workability} of defects in the Defects4J database, as defined in the previous section.
Next, we report on the problems we encountered in Section~\ref{sec:problems} and propose solutions in Section~\ref{sec:solutions}.

\subsection{Testing Environment}
\label{sec:environment}
As a basis for a systematic analysis, we  set up an environment for verifying that the defects in Defects4J meet the \emph{workability} requirements stated in Section~\ref{sec:requirements}.
We used a virtual machine running Ubuntu 22.04 and installed Defects4J version 2.0 with all its dependencies, including OpenJDK version 8.
We paid attention to all further setup requirements stated in the README of Defects4J, such as setting the correct timezone environment variable.
Additionally, we install our implementation that executes the tests on the Defects4J defects.

\begin{algorithm}
    \centering
    \begin{algorithmic}[1]
        \REQUIRE $bug\_id$ in Defects4J database
        \ENSURE Workability
        \STATE $checkout\_dir \gets defects4j.checkout(bug\_id)$
        \STATE $defects4j.compile(checkout\_dir)$
        \STATE $compilation\_classpath,\ test\_classpath,\ source\_dir,\newline test\_classes \gets defects4j.export(bug\_id)$
        \STATE $ast \gets parse(source\_dir)$
        \STATE $classes \gets compile(ast,\ compilation\_classpath)$
        \IF{compilation failed}
            \RETURN false
        \ENDIF
        \STATE $test\_result \gets runTestSuite(classes,\ test\_classes,\newline test\_classpath)$
        \FORALL{$test$ in $test\_result$}
            \STATE $single\_result \gets runSingleTestMethod(\newline classes,\ test,\ test\_classpath)$
            \IF{$single\_result \neq test\_result_{test}$}
                \RETURN false
            \ENDIF
        \ENDFOR
        \STATE $expected\_result \gets defects4j.export(bug\_id)$
        \RETURN $test\_result = expected\_result$
    \end{algorithmic}
    \caption{Setup-test for checking \emph{workability} of defects}
    \label{alg:setup-algorithm}
\end{algorithm}

Algorithm~\ref{alg:setup-algorithm} shows how we test that a defect from the dataset is \emph{workable}.
Given a bug ID, the buggy version is checked out and initially compiled with the Defects4J tooling (lines 1 and 2).
This ensures that all necessary steps of the build system such as copying resource files are correctly prepared.
In the following step, the relevant metadata (classpaths, source directory, etc.) are retrieved using the export command of the Defects4J tooling (line 3).

The source files are then parsed (verifying \textbf{R1}) and compiled (verifying \textbf{R2}) without modification (lines 4 to 8).
This is the first step where a negative result can occur: if parsing and compiling the unmodified source code fails, there is some problem in the project setup (e.g., compilation classpath wrong).
If compilation succeeds, the whole test suite is executed on the compiled (unmodified) program (line 9; verifying \textbf{R3}).
This creates a list of all tests in the test suite, including the results of the individual tests (see \textbf{R4}).

The test suite is executed with a \emph{test-driver}, a small program which invokes the JUnit API to run the tests and reports the results.
This is a typical implementation seen in APR tools: the test-driver executes the JUnit test cases and provides an interface to the main APR tool; the APR tool tells the test driver which part of the test suite to execute, and the test-driver reports the results back to it.
This setup enables the individual test result reporting described in \textbf{R4}, as the test-driver can pass each single test case result in a machine-readable form to the APR tool.
It also enables the execution of only individual tests described in \textbf{R5}, as the test-driver controls the execution via the JUnit API.
The test suite is usually executed in its own sub-process to isolate it from the main APR process, guarding against problems such as crashes and infinite loops in the test suite.

Following the execution of the whole test suite in Algorithm~\ref{alg:setup-algorithm}, each test is run individually with coverage information (see \textbf{R5}) and it is checked that its result does not differ from when it was executed in the complete test suite (lines 10 to 15; verifying \textbf{R6}).
This is the second instance where a negative result can be detected: if a test method has a different result when being executed on its own, then we consider the test suite as inconsistent.
In practice, this would be a problem for the fault localization step in program repair tools, as this typically executes test methods individually with coverage tracking.

Finally, the Defects4J tooling is once again called to get the set of expected failing tests and this is compared with our actually observed test result (lines 16 and 17; also validating \textbf{R6}).
This is the third and final negative result possibility: the actual failing tests differ from the ones documented in the Defects4J database.

This algorithm is executed on each defect in the Defects4J dataset.
For each run over all defects, we vary how many defects are tested in parallel, to check if a higher or lower overall system load causes the test suites to behave differently.
This creates another opportunity for a negative result: while a defect may be \emph{workable} when tested once, it may not be when executed multiple times or under higher system loads (validating \textbf{R6}).
We call these defects \emph{flaky}.
For this reason, we run 20 rounds of the setup-test in total, with varying levels of parallelism (from 1, i.e.\ serial tests of defects, to 25 defects being tested in parallel).

The Defects4J authors recommend in their README to not do compilation and test execution outside of the Defects4J framework, i.e.\ they want compilation and test suite execution to be done using the Defects4J command-line tool.
However, there are a few problems with using the command-line tool like this in APR tools:
\begin{itemize}
    \item
    Integrating Defects4J as a dependency for the compilation and validation of variants makes the APR tool not generically applicable to repairing other projects or defect datasets.
    As discussed in \textbf{R2}, simply invoking the compiler with the correct compilation classpath is preferred.

    \item
    Using the Defects4J tooling to compile variants and execute the test suite requires overwriting the original checkout location of the buggy program version.
    This prohibits parallelism in evaluating variants.
    It also has the potential for bugs, when not properly reverting all changes of a variant before validating the next one.

    \item
    Defects4J has a significant runtime overhead for compiling and for executing the test suite.
\end{itemize}

For these reasons, we use the export command of the Defects4J command-line tool to get the necessary metadata (line 3 in Algorithm~\ref{alg:setup-algorithm}) and do the compiler invocation and test suite execution in our own implementation.
This is also in line with most other APR tool implementations that we are aware of that were evaluated on the Defects4J dataset.
However, some APR evaluations do directly use the Defects4J command-line tool and are, because of this, unaffected by some of the problems we describe in the next section.

\subsection{Encountered Problems}
\label{sec:problems}
When executing the setup-test described in the previous section, we encountered multiple different problems.
These range from technical challenges to problems in the dataset itself.
Here, we will describe these problems in more detail, while the next section describes (sometimes only partial) proposed solutions.
Technical problems highlight aspects that implementations of evaluation frameworks for APR tools should pay attention to.
Even though they may seem less relevant from a pure research-level point of view, we hope that they may help future work to more easily overcome the technical challenges we encountered.

We cannot report accurate numbers for how many defects are affected by each individual problem, as they shadow each other (e.g.\ a compilation problem hides problems in the test suite, as it cannot be executed without a working compilation).
Also, the (partial) solutions for each problem, which are in some cases required to even conduct an investigation, change the number of affected defects for each problem.
Thus, we only report the number of non-\emph{workable} bugs in the next section, when all our proposed solutions are used.

\begin{table*}
\centering
\small
\begin{tabular}{l p{3.5cm} p{4cm} l l p{3.8cm}}
\hline
\textbf{ID} & \textbf{Problem} & \textbf{Affected approaches} & \textbf{Req.} & \textbf{Nature} & \textbf{Solution status} \\
\hline

1 & Interference between parallel executions & Parallelized benchmark executions & R3, R6 & Technical & Addressed by isolating temporary directories \\

2 & Compilation fails & APR tools compiling variants outside Defects4J tooling & R2 & Dataset/technical & Partially addressed by metadata corrections \\

3 & Test execution fails & APR tools executing tests outside Defects4J tooling & R3 & Dataset/technical & Partially addressed by metadata corrections \\

4 & JUnit version conflict & APR tools using custom test-drivers & R3--R5 & Technical & Addressed by minimizing test-driver dependencies and injecting a single JUnit version \\

5 & Classpath conflicts in test execution & APR tools using coverage/fault-localization libraries in test-driver & R3, R5 & Technical & Addressed by minimizing test-driver dependencies \\

6 & Actually failing tests differ from expected & All & R3, R6 & Dataset/reproducibility & Not addressed; exclude affected defects \\

7 & Individual test execution differs from full test suite execution & Coverage-based fault localization and APR tools using test-case prioritization & R5, R6 & Test suite quality & Not addressed; exclude affected defects \\

8 & Flaky results across repeated executions & All & R6 & Test suite quality & Mostly not addressed; exclude affected defects \\
\hline
\end{tabular}
\caption{Summary of encountered problems, affected APR approaches, related workability requirements, problem nature, and solution status.}
\label{tab:problem-summary}
\end{table*}

Table~\ref{tab:problem-summary} summarizes the problems discussed in this section.
For each problem, we indicate which kinds of APR approaches are affected, which workability requirements from Section~\ref{sec:requirements} are violated or threatened, whether the issue is primarily technical or reflects a deeper test suite or dataset problem, and whether our proposed solutions in Section~\ref{sec:solutions} address it.

\textbf{Problem 1.}
When running the setup-test on multiple defects in parallel, the test suites from the same underlying project tend to interfere with each other, because they read and write to the same temporary file locations.
While this problem could be worked around by simply not evaluating different defects in parallel, in practice this parallelization is very often used to speed up the overall runtime of an experiment.
As an illustrative example, running the setup-test described in Section~\ref{sec:environment} on all defects in Defects4J takes just over 95 hours when run serially, while it reduces to 7 hours and 45 minutes when run with 15 instances in parallel.
This effect only grows larger, the more runtime is required per individual defect, so APR approaches with high runtime are even more affected.

\textbf{Problem 2.}
For many defects in the dataset, compilation fails (line 7 in Algorithm~\ref{alg:setup-algorithm}).
This indicates a problem in the compilation classpath as exported by Defects4J tooling.
See also requirement \textbf{R2}.
This problem is immediately obvious when evaluating the defects, as a failed initial compilation prohibits any further experimentation on a given defect.
With the initial setup of our evaluation framework without any of the proposed solutions from next section, the number of non-compiling defects was 244, almost 30\,\% of the dataset.
This large number motivated us to look deeper into the underlying technical challenges of working with the dataset, as this amount of problematic defects cannot simply be ignored.

Note that, when using the Defects4J command-line tool, all of these instances do correctly compile, which hints at the exported classpath and the one used internally by the command-line tool being different.
This also means that APR evaluations which use the Defects4J tool directly are unaffected by this problem.
However, as we argued in Section~\ref{sec:environment}, not using the Defects4J tool for compilation has significant advantages, thus it is still worth to address this problem.

\textbf{Problem 3.}
For some defects in the dataset, execution of the test suite fails because dependencies of it are not found.
This indicates a problem in the test classpath as exported by Defects4J tooling.
See also requirement \textbf{R3}.
This problem is, in essence, similar to \textbf{Problem 2}, but instead of the compilation classpath, the test classpath is problematic.
In this context, we noticed obviously erroneous entries in the test classpaths output by the Defects4J tool, such as the literal strings \texttt{file} and \texttt{\$\{test.classes.dir\}} in the defects from the Cli project.
This indicates problems in the Defects4J tool implementation, as it seems to show missing variable expansions.

\textbf{Problem 4.}
Even when the test classpath is correct, it typically contains the JUnit library, which may conflict with the JUnit version included in the test-driver of the evaluation framework.
The practical need for a test-driver in APR tools and our evaluation framework is described in Section~\ref{sec:environment}.
Implementing a test-driver without a JUnit dependency is not possible, as it needs to use the JUnit API to be able to only run specific single test methods (\textbf{R5}) and collect their result in a systematic fashion (\textbf{R4}).

\textbf{Problem 5.}
We found that in practice other APR tools include additional JARs in the test classpath, often because their test-driver depends on them.
As an example, in the FLACOCO tool for fault localization~\cite{SilvaMartinezDanglot+21}, the test-driver does the coverage analysis in the sub-process.
This requires the JaCoCo library for coverage tracking and its dependencies to be included in the test-driver classpath.
In the defects from the Mockito project in Defects4J, this causes a conflict as their test classpath contain a different, binary incompatible version of JaCoCo.
This causes the test suite execution to fail in different ways, depending on the order in which the dependencies are loaded by the Java virtual machine.

In other tools we found that the test classpath used to invoke the test suite is loaded with additional JARs, not specified by the Defects4J tool.
We suspect that this done to circumvent Problem 3; the classpath is expanded to make more test suites run.
Simply adding a bunch of JARs to the classpath can be problematic, as different (incompatible) versions of the same library can cause the test suite to behave differently, as described in the case of FLACOCO.
As with the erroneous entries in the classpaths output by Defects4J itself (described in Problem 3), there seems to be a lack of care when working with classpaths, as simply adding more entries to it does not create immediately obvious problems.

\textbf{Problem 6.}
For some defects in the dataset, the actual failing tests differ from the ones expected as described in the Defects4J database (line 17 in Algorithm~\ref{alg:setup-algorithm}).
This of course is problematic, as a defect should be indicated by the test suite in a reproducible manner.
Note that, as with \textbf{Problem 2}, this problem does not occur when using the Defects4J command-line tool to execute the tests.

\textbf{Problem 7.}
For many defects in the dataset, the results of individual tests differ when executed individually with coverage versus when executing the whole test suite (line 13 in Algorithm~\ref{alg:setup-algorithm}).
See also requirements \textbf{R5} and \textbf{R6}.
This leads typical fault-localization techniques that require per-test coverage information to work on different test results, which causes its result to be less accurate or even wrong.
Also affected are techniques that reduce the number of tests being executed per evaluation during the APR approach, such as only re-running a subset of relevant test cases.
Both, the fault localization and the reduction of the test suite to a relevant subset of tests have the implicit assumption that tests behave the same, whether executed in the whole test suite or individually.
We find here that this assumption does not hold for all defects.

\textbf{Problem 8.}
For some defects in the dataset, the initial execution of the setup-test reports that they are \emph{workable}, but in further executions (also with varying system loads) they become flaky and sometimes fail.
In the worst case, a defect was identified as flaky only after 13 executions of the setup-test.
See also requirement \textbf{R6}.

\subsection{Proposed Solutions}
\label{sec:solutions}
We attempted to resolve the identified problems for the whole Defects4J dataset.
This was on the one hand driven by our own needs for using it in APR and on the other hand by an attempt to truly understand the limits of the \emph{workability} for this set of defects.

Our solution for \textbf{Problem 1} is to implement separated temporary directories for the test suite executions in our evaluation framework.
Technically this is done by setting a property for the Java virtual machine running the test suite.
Proper separation of test suite executions like this enables parallelism, drastically reducing the runtime of experiments running on the whole dataset.

Our solution for \textbf{Problem 2} and \textbf{Problem 3} is, based on manual investigation, to apply corrections to the metadata provided by the Defects4J command-line tool.
For the projects Cli, Closure, JacksonXml, Mockito, Math, and Time the compilation and/or test classpath have to be adapted by adding or removing entries\footnote{Implemented in the class \texttt{Defects4jWrapper} in our evaluation framework, see Section~\ref{sec:data}}; otherwise compiling variants or executing the test suite would fail for some or all defects in these projects.
Further, the Lang project needs to be parsed using the ISO 8859-1 encoding, while all other projects require the UTF-8 encoding; the required encoding is not specified by Defects4J.
This reduces the number of outright compilation and parsing problems to 26.
Note that even with the manual investigation, we were unable to get all defects in the dataset to compile.
All but 2 of these remaining ones are the defects 42 to 65 from the Lang project; this indicates that there is some peculiarity in the build system of the Lang project in the timeframe that these defects stem from.

To address \textbf{Problem 4} and \textbf{Problem 5}, we designed the test-driver in our evaluation framework to be as small as possible.
Its only dependency is JUnit, as it needs this library to execute tests.
Luckily, JUnit has a very good binary compatibility across  versions 3.8.x and 4 (these versions cover all test suites in the Defects4J dataset), so we can remove the JUnit JARs from the test classpath reported by Defects4J and inject our own version of it instead.
Thus our implementation does not introduce any dependency conflicts, like the ones described in \textbf{Problem 5}.

\begin{table}
    \centering
    \begin{tabular}{l l}
        \textbf{Bug ID(s)}  & \textbf{Exclusion Reason} \\
        \hline
        Chart/5--26 & Inconsistent test suite \\
        Cli/7 & Inconsistent test suite \\
        Cli/13--16 & Inconsistent test suite \\
        Cli/21 & Inconsistent test suite \\
        Codec/17--18 & Result differs from dataset \\
        Collections/25 & Inconsistent test suite \\
        Collections/26--27 & Compilation failed \\
        Collections/28 & Inconsistent test suite \\
        Compress/30--47 & Inconsistent test suite \\
        JacksonCore/19 & Result differs from dataset \\
        JacksonCore/25 & Result differs from dataset \\
        JacksonDatabind/1--6 & Flaky \\
        JacksonDatabind/95 & Flaky \\
        Jsoup/67--93 & Inconsistent test suite \\
        Lang/1 & Inconsistent test suite \\
        Lang/3--4 & Inconsistent test suite \\
        Lang/42--65& Compilation failed \\
        Math/1--28 & Inconsistent test suite \\
        Math/29--37 & Flaky \\
        Math/40 & Flaky \\
        Math/53--54 & Flaky \\
        Math/59 & Flaky \\
        Math/62--63 & Flaky \\
        Math/70 & Flaky \\
        Mockito/1--11 & Inconsistent test suite \\
        Mockito/18--26 & Inconsistent test suite \\
        Time/27 & Result differs from dataset \\
        \hline
    \end{tabular}
    \caption{Non-\emph{workable} defects in Defects4J 2.0}
    \label{tab:exclusions}
\end{table}

For \textbf{Problem 6}, \textbf{Problem 7}, and \textbf{Problem 8} we could  not find a good solution.
We could only partially improve the situation regarding \textbf{Problem 8} in the Math project, by removing the test class FixedElapsedTimeTest from the test suite; it depends on timing measurements during execution and thus is very susceptible to varying system loads.
All other problematic defects cannot be addressed and thus we recommend to exclude them from any APR experiments done using Defects4J.
Notably, this includes all four defects from the Collections project, effectively reducing the \emph{workable} subset of Defects4J to only 16 real-world Java projects.
Table~\ref{tab:exclusions} shows all defects that should be excluded from future APR experiments, as well as the reason for exclusion, for reference.

Particularly concerning is the range of flaky defects in the Math project (Math/29 through Math/70).
While the defects Math/29 through Math/37 are failing quite often (only passing the setup-test 1 or 2 times in the 20 total executions), the other flaky ones only present themselves very sporadically (most of them fail the setup-test only once during the 20 executions).
The danger here is that they seem to be \emph{workable} most of the time, and significant effort is required to reveal them as flaky.
Investigating which specific tests in the test suite are flaky reveals some different tests for different defect instances, but with the test \texttt{testWeightedConsistency} in the \texttt{MeanTest} class occurring as flaky multiple times in different defect instances.
This test is present in the same way in many other defect instances of the Math project, which have not been found to be flaky in our series of setup-tests.
Thus it raises the question, whether more defect instances in this range of Math project versions are flaky, but have not yet been found because the probability of them presenting themselves as flaky is very low.

\begin{figure}
    \centering
    \definecolor{pastelred}{RGB}{255, 179, 186}  
    \definecolor{pastelorange}{RGB}{255, 204, 153}  
    \definecolor{pastelpurple}{RGB}{200, 150, 255}  
    \definecolor{pastelbrown}{RGB}{210, 180, 140}  
    \definecolor{pastelgreen}{RGB}{119, 221, 119}   
    \begin{tikzpicture}
        \pie[text=legend,radius=2,color={pastelgreen,pastelred},sum=auto,change direction]{
            666/\emph{Workable},
            180/Not \emph{Workable}
        }
    \end{tikzpicture}
    \caption{Ratio of \emph{workable} defects in Defects4J 2.0.0}
    \Description{A pie chart showing the relative amount of workable and non-workable defects.}
    \label{fig:reproducible-ratio}
\end{figure}
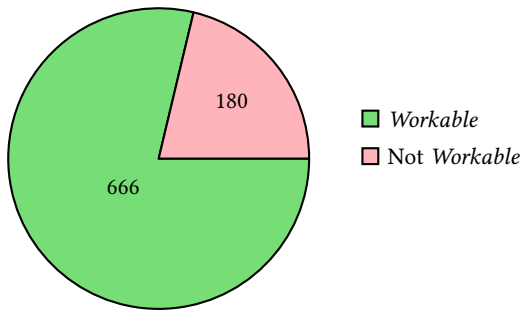

\begin{figure}
    \centering
    \definecolor{pastelred}{RGB}{255, 179, 186}  
    \definecolor{pastelorange}{RGB}{255, 204, 153}  
    \definecolor{pastelpurple}{RGB}{200, 150, 255}  
    \definecolor{pastelbrown}{RGB}{210, 180, 140}  
    \definecolor{pastelgreen}{RGB}{119, 221, 119}   
    \begin{tikzpicture}
        \pie[text=legend,radius=1.9,color={pastelred, pastelorange, pastelpurple, pastelbrown, pastelgreen},sum=auto,align=left]{
            26/Compilation fails,
            126/Inconsistent test suite,
            5/{Result differs \\[-1ex] from dataset},
            23/Flaky
        }
    \end{tikzpicture}
    \caption{Exclusion reasons of non-\emph{workable} defects in Defects4J 2.0.0}
    \Description{A pie chart showing the relative amount of non-workability reasons: compilation fails, inconsistent test suite, result differs from dataset, flaky.}
    \label{fig:problem-type-ratio}
\end{figure}
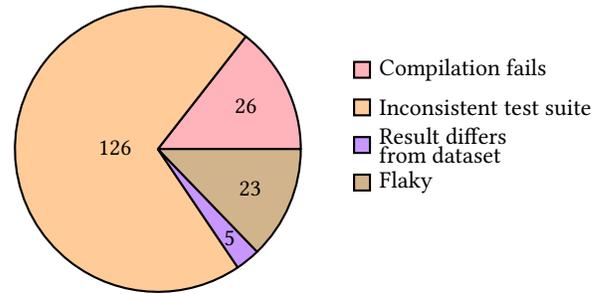

Figure~\ref{fig:reproducible-ratio} shows the ratio of \emph{workable} defects in the Defects4J dataset, i.e., how many defects passed and how many failed the setup-test described in Section~\ref{sec:environment}.
We can see that only 78.4\,\% of the defects in the dataset are actually \emph{workable} in the context of APR evaluations.
Figure~\ref{fig:problem-type-ratio} shows for the 180 defects that are not \emph{workable}, the reason why they are excluded.
The categories correspond to the negative results returned by Algorithm~\ref{alg:setup-algorithm} as well as the \emph{flaky} judgment if results differ in any of the repeated executions.
Note that the order we check the different exclusion reasons matters, as the ones checked earlier in Algorithm~\ref{alg:setup-algorithm} shadow the ones checked later.
For example, if we find an inconsistent test suite (return statement in line 13), we do not check the test result against the one documented in the Defects4J dataset (line 17).

The largest reason for exclusion is that running single tests individually results in different results than running the test suite as a whole (``inconsistent test suite''  in Figure~\ref{fig:problem-type-ratio}).
This is especially concerning, as this inconsistency is not obvious when not specifically testing for this; an APR tool that does not compare the individual test results during fault localization with the ones from the test suite as a whole will be oblivious to this problem, and thus base its process on inconsistent data.
It also does not only affect fault localization, but also further evaluations of variants generated during the APR process itself; if single tests behave differently when run out-of-order, the results of evaluations on only parts of the test suite (which are very common in APR tools) cannot be trusted anymore.
Similarly concerning are the 23 defects we found to be flaky, as it indicates more subtle problems in the test suite that only show up randomly.
In contrast, the 5 defects where the actual test result differs from the one documented in the Defects4J dataset (``result differs from dataset'' in Figure~\ref{fig:problem-type-ratio}) can easily be checked for.

The 26 defects that do not even compile (``compilation fails'' in Figure~\ref{fig:problem-type-ratio}) indicate remaining technical challenges in the reproducibility of the defects.
The compilation does succeed when invoking the compile command in the Defects4J command-line tool itself, so there is a difference between the compilation setup between the exported classpath and the Defects4J framework.
In contrast to repeatability, which stipulates same results in the same setup, reproducibility requires same result in different setups as well.
As discussed earlier, it is common for implementations of APR techniques to do compilations and test suite evaluations in a custom setup, rather than using the Defects4J framework for these tasks.
On the bright side, a failing compilation is a rather obvious problem that will likely not go unnoticed and thus has no hidden effect on evaluation results.

\section{Quality of Test-Suites}
\label{sec:test suite-quality}

We investigate how well the test suites provided by the projects included in the Defects4J dataset are able to discriminate incorrect program variants, i.e.\ variants that are not proper fixes for the underlying defect.
In essence, this represents an even stricter requirement on the defects in the dataset than the \emph{workability} we have looked at in the previous sections.
The developers of the underlying projects who created the test suites did not create them with the usage for APR tools in mind, so there is a risk that the test suites are not adequate for the APR setting.

Many APR approaches generate a lot of program variants and the test suite is used to validate if they fix the defect.
Ideally, the test suite only lets correct variants for the underlying defect pass.
However, in practice there is a distinction between \emph{plausible} patches (that pass the test suite) and \emph{correct} patches (which changes the implementation to exhibit the intended behavior)~\cite{LongRinard16}.
In APR experiments, this difference is sometimes investigated by manual inspection.
As correct patches are hard to identify, the automatically created patch is compared to the the human-written one in these manual inspections.
In other experiments, only the plausible fixes are reported and no manual investigation for correctness is done; such results build on the ability of the test suites to not let incorrect patches pass.
We investigate this ability in the following experiment.

We experimentally check how well the test suites are able to reject a certain kind of small, usually nonsensical patch: single statement deletion.
The reasoning here is that especially search-based approaches do small, iterative modifications to the program at random.
Many of these variants are nonsensical and thus should be rejected by the test suite.
In our work on search-based APR tools, we found that especially the statement-deletion operator is surprisingly prominent in variants that are found to make the test suite pass.
With our experiment, we investigate how well the test suites in the defect dataset are able to root out these kind of incorrect patches.

We conduct an experiment where we iterate through the buggy program, and one-by-one delete a single statement and check if the test suite passes with this modification.
For this, we implemented fault localization similar to the FLACOCO tool \cite{SilvaMartinezDanglot+21}, which does spectrum-based fault localization utilizing the Ochiai coefficient~\cite{AbreuZoeteweijVanGemund06}.
Candidates for deletion are statements with a suspiciousness value of at least 0.01 and limited to 300 statements in total (a limitation to remain within reasonable computational effort).
Additionally, a maximum runtime of 3 hours per defect is enforced by terminating the process if this timeout is reached.

The result is, that 69 defects out of the 655 \emph{workable} ones in Defects4J are ``fixed'' by deleting just a single statement in this manner, which would correspond to a fix-rate of 10.5\,\%.
We further manually inspect the human-written patches for all of these defects and find that in only 10 of these cases they only consist of deletions (incl.\ cases where multiple statements are deleted).
This means that for 59 defects (9.0\,\% of all \emph{workable} defects) just deleting a single statement is very likely not a correct patch, but nevertheless the test suite passes in this case.

\begin{figure}
    \centering
    \begin{subfigure}{\columnwidth}
        \centering
        \begin{lstlisting}[language=diff]
--- a/org/apache/commons/cli/HelpFormatter.java
+++ b/org/apache/commons/cli/HelpFormatter.java
@@ -629,7 +629,7 @@ public class HelpFormatter {
   // if the Option has a value
-  if (option.hasArg() && (option.getArgName() != null))
+  if (option.hasArg() && option.hasArgName())
   {
     buff.append(" <")
         .append(option.getArgName())
         .append(">");
   }
        \end{lstlisting}
        \vspace{-10pt}
        \caption{Human-written correct patch}
        \label{fig:comparison-human}
    \end{subfigure}
    \begin{subfigure}{\columnwidth}
        \vspace{10pt}
        \centering
        \begin{lstlisting}[language=diff]
--- a/org/apache/commons/cli/HelpFormatter.java
+++ b/org/apache/commons/cli/HelpFormatter.java
@@ -631,7 +631,4 @@ public class HelpFormatter {
   // if the Option has a value
   if (option.hasArg() && (option.getArgName() != null))
   {
-    buff.append(" <")
-        .append(option.getArgName())
-        .append(">");
   }
        \end{lstlisting}
        \vspace{-10pt}
        \caption{Automatically generated patch}
        \label{fig:comparison-auto}
    \end{subfigure}
    \caption{Comparison of human-written and automatically generated patch (some formatting changes for readability)}
    \Description{Two patches comparing the human-written and automatically generated patches for the Cli/11 defect in Defects 4j. The human-written one replaces a condition of an if, while the automatically generated one deletes the only statement enclosed by the if block.}
\end{figure}

One example defect from Defects4J that is incorrectly passing the test suite when deleting a single line is Cli/11.
Figure~\ref{fig:comparison-auto} shows the patch when deleting just a single line, which makes the test suite pass.
We actually encountered this ``fix'' being produced by a search-based APR tool when run on this defect.
Figure~\ref{fig:comparison-human} shows the actually correct patch, which was written by a human.
We see that the correct patch is to change the condition from a simple null-check to a call to \texttt{hasArgName} (which also checks for an empty string).
However, there is no test in the test suite that checks whether the content in the if-body works correctly.
Thus, simply deleting the contents of the if-body, instead of making the condition more restrictive, is a valid patch according to the test suite.
This example highlights how an underspecified test suite can be ``gamed'' by an APR tool.

\begin{table}
    \centering
    \begin{tabular}{|l|c|c|c|}
        \hline
        \textbf{Data(sub)set}       & \textbf{\#Defects} & \textbf{\#Fixed} & \textbf{Fix-Rate} \\ \hline
        Original Defects4J 1.0      & 357                & 49               & 13.7\,\%          \\ \hline
        excl.\ not in Defects4J 2.0 & 353                & 49               & 13.9\,\%          \\ \hline
        excl.\ non-\emph{workable}  & 259                & 31               & 12.0\,\%          \\ \hline
        excl.\ trivial              & 239                & 19               & 7.9\,\%           \\ \hline
    \end{tabular}
    \caption{Comparison of number of defects fixed in jGenProg evaluation for \emph{workable} and non-trivial subsets}
    \label{tab:jgenprog}
\end{table}

The amount of defects we consider non-\emph{workable} (Section~\ref{sec:defects4j}) or have too-easy-to-fix test suites has a real impact on empirical evaluations of APR tools that use Defects4J.
As an illustrative example, we analyze the potential impact on the evaluation results of the search-based APR tool jGenProg~\cite{MartinezMonperrus19}.
Table~\ref{tab:jgenprog} summarizes the fix-ratios when excluding non-\emph{workable} and trivial-to-fix defects.
The evaluation of jGenProg uses Defects4J version 1.0, which has 357 defects, and finds that 49 of these are fixable by this repair approach~\cite{MartinezMonperrus19}.
From Defects4J version 1.0 to 2.0, 4 defects were excluded because they do not reproduce with the updated Java version (Defects4J 2.0 switched from Java version 7 to 8).
The second row in Table~\ref{tab:jgenprog} shows the remaining set; none of the excluded defects were fixed by jGenProg, thus this reduction slightly increases the fix-rate.
Of all these defects both in version 1.0 and 2.0, 94 are considered non-\emph{workable} by us (see Section~\ref{sec:solutions}).
The third row shows that, when removing the deprecated and non-\emph{workable} defects, 259 defects remain, of which jGenProg can fix 31.
We then further remove 20 defects from this subset, which we found fixable with a single-statement deletion, although their human-written patch does not just delete code.
The result of this is shown in row four: of the remaining 239 defects, jGenProg can only fix 19.
Notice how the exclusion of trivially fixable defects reduces the fix-rate to less than two-thirds of the original, a much larger impact compared to the exclusion of non-\emph{workable} defects.
The problem of trivial patches that just remove functionality was also found by the authors of jGenProg during manual analysis of the patch correctness~\cite{MartinezDurieuxSommerard+17}, although they do not specifically report whether jGenProg produced non-trivial fixes for these defects.

These results show that analyzing the \emph{correctness} of patches produced by APR tools is of high importance.
The test suites included in the defect datasets cannot be relied upon to correctly identify plausible but incorrect patches.
This is also in line with previous works investigating the correctness vs. plausibility of APR patches~\cite{LongRinard16}.
Fundamentally, the test suites for the defects were not specifically created for an APR setting, and thus their ability to root out randomly created incorrect patches is not specifically engineered into them.

\section{Threats to Validity}
\label{sec:threats}
We discuss the threats to validity for our two experiments described in Section~\ref{sec:environment} (\emph{workability} of defects) and Section~\ref{sec:test suite-quality} (trivially fixable test suites) separately in the following sub-sections.

\subsection{Workability Experiment}
The \textbf{conclusion validity} of the experiment is threatened by potentially unreliable measures.
Specifically, we see with the flaky defects that certain outcomes do not reliably occur in our tests of the setup.
To some extent, we  uncovered unreliable test suites, but we cannot be sure that we found all.
Thus, the number of non-\emph{workable} defects in the dataset could actually be higher.

The \textbf{internal validity} is threatened by possible mistakes in our implementation of the evaluation framework.
We paid special attention to create a robust framework that tries to detect problematic or inconsistent situations and brings them to our attention by failing in these cases.
However, we cannot be completely sure that all problems will always be detected.
Further, we cannot be sure that specifics of our experimental setup may have caused the problems we observed, although we paid attention to follow all documentation on the required environment by the Defects4J authors and performed significant experimentation to ensure the reliability of the setup.

The \textbf{construct validity} of the experiment is threatened by the way we test for inconsistent test suites.
Other methods for detecting inconsistent test results could be used, such as explicitly randomizing test method execution order, or running each test method in its own Java virtual machine, instead of reusing the same process, etc.
Also, manual inspection of the test suites could be done to detect dependencies between test cases.
This would lead to greater insight as to why the test suites are inconsistent, though this requires great manual effort.
On the other hand, the way we detect inconsistent test suites is, as discussed earlier, very close to how automated program repair tools execute the test suites in practice.
Thus, we argue that the problems we detect with our measurement approach are very close to those of practical importance.

As for  \textbf{external validity}, we only analyzed a single defect dataset (Defects4J version 2.0) and make no claims as to the quality of other datasets.
Still, the external validity is threatened by our evaluation setup: the observed effects could have only occurred due to our specific setup.
However, in the creation of our setup we were guided by the evaluation setups we have seen in actual evaluations of automated program repair approaches, as well as by the documentation of the Defects4J authors on how to set up an evaluation environment.
Thus, we argue that the setup is in the relevant aspects similar to (and thus all observations we made will also occur in) all other environments that use Defects4J as an evaluation dataset.

\subsection{Test-suite Quality Experiment}
The \textbf{conclusion validity} of the experiment is threatened by possible random fluctuations in the measurements, i.e.\ the test suite result.
We only ran the experiment on the defects we deemed \emph{workable} in Section~\ref{sec:defects4j}, which already removes all test suites we know to be flaky.
Additionally, we repeated the experiment three times, and always got the exact same result.

The \textbf{internal validity} of the experiment is threatened by the way we select the deletion candidates.
We do not consider every single statement in the program for deletion, but only those statements above a certain suspiciousness threshold and also limit it to a certain maximum total number of statements.
It could be that further defects can be fixed by deleting only a single statement, which we did not find because of this.
However, this would mean that the actual number of too-trivial-to-fix defects might be even higher than what we find, which would only strengthen our argument.

The \textbf{construct validity} is threatened by the way we generate the small, nonsensical patches: deleting only a single statement.
Further possibilities would be to also consider combinations of multiple statements to delete as well as replacement of statements with trivial alternatives (e.g.\ ``return~0'').
This would, however, significantly increase the search space and thus runtime of the experiment.
The effect of this threat is also like for the internal validity one described earlier that the actual number of too-trivial-to-fix defects might be actually higher  than what we found.

As for \textbf{external validity}, the same reasoning as for the \emph{workability} experiment discussed in the previous sub-section applies.
In general, we also only make claims about the specific Defects4J version 2.0 dataset, and not about other ones, although given the overall rather positive analysis of Defects4J in other sources~\cite{ZhuFurthPradel+25, ZhuRubioGonzalez23} we can expect that in those problems might be even more significant.

\section{Conclusion}
\label{sec:conclusion}
In this paper, we took a critical look at the most commonly used evaluation dataset of defects for automated program repair (APR) approaches: Defects4J.
We analyzed both technical and more fundamental challenges, which occur when trying to work with this dataset in practice to evaluate APR tools.
For this, we introduced in Section~\ref{sec:requirements} the requirements that a defect in the dataset should fulfill to be \emph{workable} for evaluations of APR approaches.
We contrasted this with  actual problems we encountered  in Section~\ref{sec:problems}.
Our analysis is based on a systematic evaluation of all defects in the dataset (Section~\ref{sec:environment}).
This resulted in several observations.
Only 78.4\,\% of the defects are \emph{workable}, most non-\emph{workable} are because the test suite returns inconsistent results when running tests individually.
Our main contributions are guidelines that provide solutions for the encountered problems as well as an evaluation framework that implements the solutions to the technical problems and can be used as a basis for future evaluations of APR tools and defect datasets.

Further, in a second experiment, we investigated the quality of the underlying test suites of the defects to determine potentially incorrect patches (Section~\ref{sec:test suite-quality}).
Here, we found a significant number of under-specified test suites (9.0\,\% of all \emph{workable} defects), which is in line with other literature on this topic~\cite{MartinezDurieuxSommerard+17}.
We also found that excluding these too-easy-to-fix defects from the dataset would have significant influence on past evaluations that used Defects4J, which demonstrates the importance for future research that uses Defects4J to take our results into account.

\section{Data Availability}
\label{sec:data}
Our implementation of the evaluation framework and setup-test algorithm introduced in Section~\ref{sec:environment} as well as the detailed results from the setup-test and the statement-deletion (Section~\ref{sec:test suite-quality}) experiments are available on Zenodo: \url{https://doi.org/10.5281/zenodo.18834780}

\bibliography{bibliography}

\end{document}